\documentclass[cits]{PoS}

\pdfoutput=1 

\usepackage{url}
\usepackage[round]{natbib}

\bibpunct{[}{]}{,}{n}{}{;}

\def\fdg{\hbox{$.\!\!^\circ$}}
\def\farcm{\hbox{$.\mkern-4mu^\prime$}}
\def\farcs{\hbox{$.\!\!^{\prime\prime}$}}

\newcommand{\arcsec}{$''$}
\newcommand{\arcmin}{$'$}
\renewcommand{\deg}{$^\circ$}

\newcommand{\hlinkbibcode}[1]{http://adsabs.harvard.edu/cgi-bin/bib_query?#1}

\newcommand{\myurl}[1]{\protect\href{#1}{\protect\url{#1}}}

\newcommand{\difmap}{{\tt difmap}}
\newcommand{\aips}{{\tt AIPS}}

\title{Long baseline experiments with LOFAR}

\ShortTitle{Long baseline LOFAR}

\author{\speaker{Olaf Wucknitz}%
  \thanks{This work is supported by the Emmy-Noether-Programme of the
    `Deutsche Forschungsgemeinschaft', reference WU\,588/1-1, and by 
    a Marie Curie European Reintegration Grant
    within the 7th European Community Framework Programme, Contract No.\
    PERG02-GA-2007-224897 `WIDEMAP'.}\\
  Argelander-Institut f\"ur Astronomie, Auf dem H\"ugel 71, 53121
  Bonn, Germany\\
  E-mail: \email{wucknitz@astro.uni-bonn.de}}

\abstract{I present first results of LOFAR observations with international
  baselines. An important cornerstone was the detection of the first
  long-baseline fringes. Their analysis turns out to be extremely useful to
  investigate and solve a number of technical issues of the
  instrument. Crude maps of the sky are created from single-baseline
  delay/fringe-rate spectra and compared with a short-baseline synthesis
  map. First long-baseline LBA images are shown of the source 3C196, a bright
  quasar whose sub-components can only be resolved with the long
  baselines. The corresponding sub-arcsec HBA image does not show the same
  amount of details yet, but HBA results are expected to improve significantly
  very soon.
 The LBA long-baseline image of 3C196 comprises the highest-resolution radio
 map ever produced at this low frequency.}

\FullConference{ISKAF2010 Science Meeting - ISKAF2010\\
		June 10--14, 2010\\
		Assen, the Netherlands}

\begin{document}

\section{Introduction}

The Low Frequency Array (LOFAR) will soon consist of 30--40 stations in the
Netherlands plus international stations in Germany, France, England, Sweden
and hopefully also Italy, Poland, Spain and other countries.
Currently the only available international stations are all in Germany:
Effelsberg near Bonn, 
Tautenburg near Jena, Unterweilenbach near Munich and Bornim near Potsdam.

LOFAR will offer wide fields of view, observations with several independent
beams (in different directions) and a high survey speed.
Two subsystems cover a frequency range between about 10 and 250\,MHz. The Low
Band Antennas (LBA) are used for 10--80\,MHz (only 30--80 will be used
regularly because of strong RFI at lower frequencies), and the High Band
Antennas (HBA) for 110--250\,MHz. The FM band is excluded intentionally.

On long baselines, sub-arcsec resolution will be possible even in the LBA band,
which opens an entirely new parameter space for radio observations. Because of
the high number of stations, full synthesis imaging is possible.

\section{Resolution}

The main motivation for international baselines is the possibility to reach
much higher resolutions than with Dutch stations alone. One exciting
application of LOFAR surveys is the search for gravitational lenses
\citep{wucknitz08}. These objects can
be identified directly from the survey data provided that sub-arcsec
resolution can be achieved. The reliability of this identification and thus
the efficiency of lens surveys depends strongly on the resolution and the
sensitivity on long baselines.
The resolution that can be achieved with LOFAR is shown in Tab.~\ref{tab:res}
as a function of frequency and baseline length.

\begin{table}[hb]
\center
\begin{tabular}{c|c|c|c|c|c}
freq/MHz & $\lambda/$m & 1\,km & 30\,km & 300\,km & 1000\,km \\ \hline
10 & 30 & 1\fdg7 & 3\farcm4 & 21\arcsec & 6\farcs2 \\[0.5ex]
30 & 10 & 34\arcmin & 1\farcm1 & 6\farcs9 & 2\farcs1 \\[0.5ex]
80 &  3.8 & 13\arcmin & 26\arcsec & 2\farcs6 & 0\farcs77 \\[0.5ex]
120 & 2.5 & 8\farcm6 & 17\arcsec & 1\farcs7 & 0\farcs52 \\[0.5ex]
160 & 1.9 & 6\farcm4 & 13\arcsec & 1\farcs3 & 0\farcs39 \\[0.5ex]
220 & 1.5 & 4\farcm7 & 9\farcs4 & 0\farcs94 & 0\farcs28
\end{tabular}
\caption{Fringe-spacing $\theta=\lambda/L$ as estimate for the resolution
  achievable with LOFAR for different combinations of frequency and baseline
  length. Sub-arcsec resolution requires baselines of at least several hundred
(HBA) or thousand (LBA) km length.}
\label{tab:res}
\end{table}

\section{International stations available for this work}

For the observations described here, only the international stations in
Effelsberg, Tautenburg and Unterweilenbach were used (see Fig.~\ref{fig:german
  stations}). Bornim was still waiting
for its network connection, and stations in other countries were not
constructed yet.

\begin{figure}
\center
\includegraphics[width=0.5\textwidth]{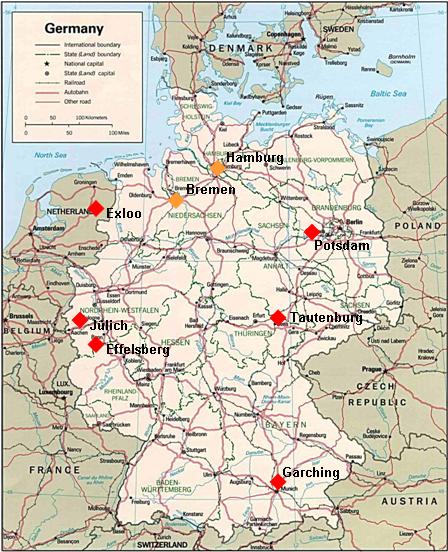}
\hspace{0.05\textwidth}
\includegraphics[width=0.2\textwidth]{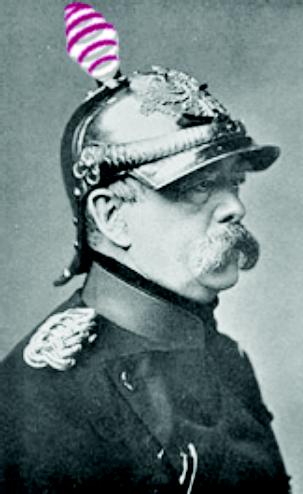}
\caption{Location of LOFAR stations in Germany (image courtesy of AIP) and the
  inofficial logo of the German subarray (`Otto-von-Bismarck-Telescope',
  invented by a colleague from Dwingeloo who is not known to be of German
  persuasion).} 
\label{fig:otto}
\end{figure}

The baseline lengths between the Dutch core and German stations are about
300\,km (Effelsberg), 400\,km (Tautenburg) and 600\,km (Unterweilenbach),
compare Fig.~\ref{fig:otto}.

\begin{figure}
\includegraphics[height=0.235\textwidth]{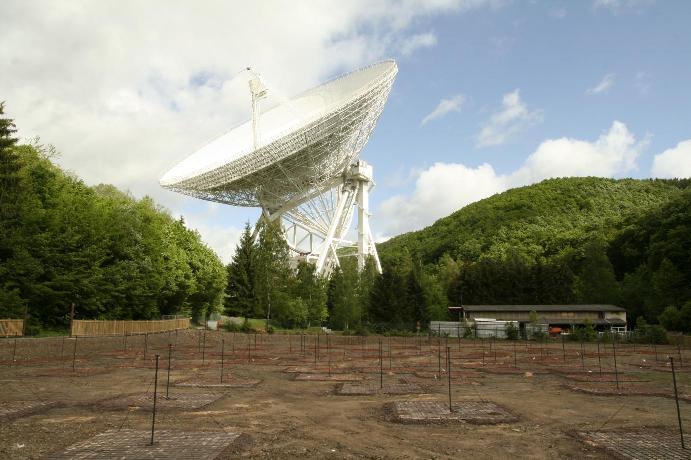}\hfill
\includegraphics[height=0.235\textwidth]{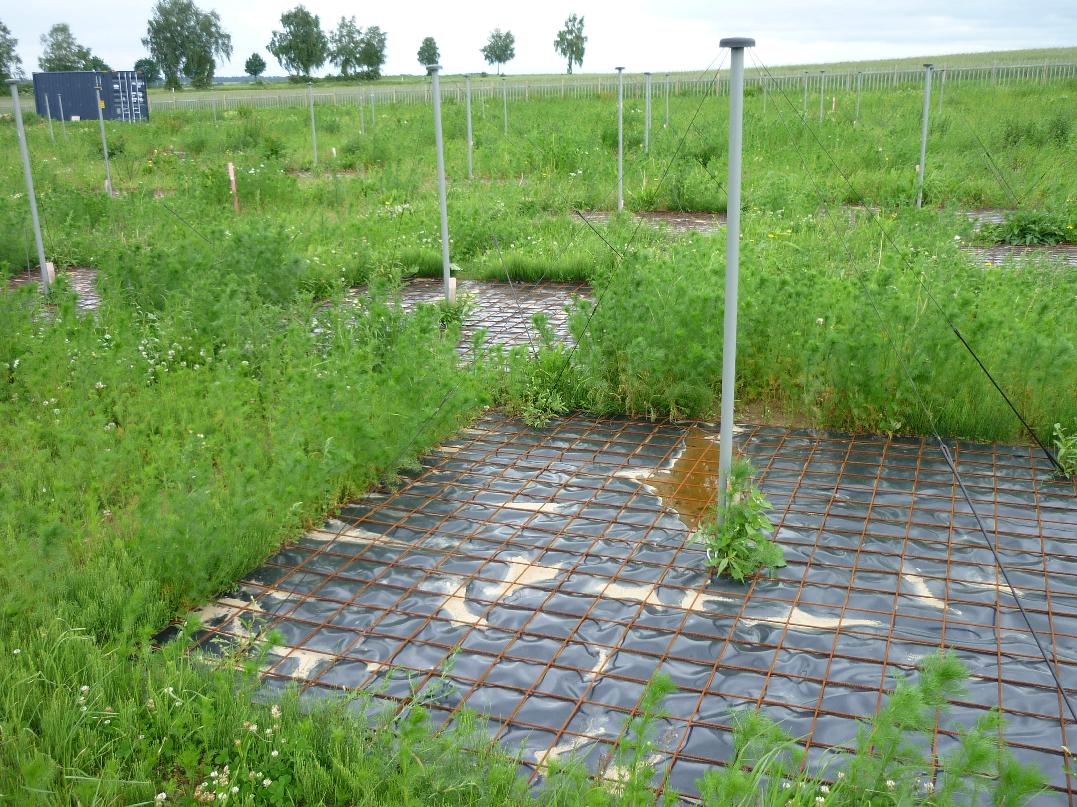}\hfill
\includegraphics[height=0.235\textwidth]{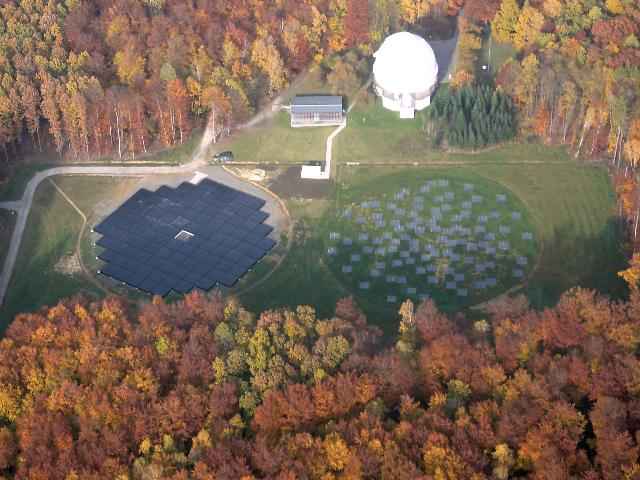}
\caption{Left: LBA field of the Effelsberg station with the 100\,m radio
  telescope   in the background, centre: LBA antennas in Unterweilenbach,
  right: the Tautenburg station with HBA and LBA fields and the dome of the
  2\,m optical telescope.}
\label{fig:german stations}
\end{figure}

\section{Long baseline issues}

A major issue on long baselines is the fact that the majority of sources will
at least be partially resolved so that the signal weakens considerably on
longer baselines. This is also true for calibrator sources so that the number
of good calibrators is reduced.

In order to achieve a sufficient signal-to-noise ratio ($S/N$) not only for imaging,
but in the first instance for the calibration, the data have to be (implicitly
or explicitly) averaged in time and frequency. In order to do that, the
dependence of phases on time and frequency has to be taken into account.
Delays (proportional to derivatives of phase with respect to frequency) and
rates (derivatives of phase with respect to time) have to be fitted and
corrected for. In Very Long Baseline Interferometry (VLBI), this technique is
known as fringe-fitting and is applied on a regular basis.

In the context of LOFAR, with the very low frequencies and very large
fractional bandwidth, the procedure is considerably more complicated than in
standard VLBI observations, because one has to distinguish between
non-dispersive and dispersive delays. Non-dispersive delays ($\tau$
independent of $\nu$) are due to clock
offsets and errors in positions of stations and sources, while dispersive
delays ($\tau\propto \nu^{-2}$) are caused by the ionosphere. Generally it
turns out that both are important. We find that the clock offsets can be much
larger than the ionospheric delays (sometimes many $\mu$sec when the clock
synchronisation fails), but they tend to be constant with time and
are thus easier to correct. Ionospheric delays of several $100\,$nsec are often
observed on long baselines at frequencies around 50\,MHz. Both effects have to
be taken into account when averaging over several subbands.

At the moment, fringe-fitting is only partly implemented in own software. I
can fit both kinds of delays on individual baselines. Because the dispersion
is not significant over a single subband, FFTs can be used within the
subbands. Interpolated results of that are combined coherently over all
subbands to determine multi-band coherent delay solutions.

With the wide bands, the rates have to be treated as delay rates instead
of phase rates, because the latter would be frequency dependent both for
dispersive and non-dispersive delays. The software thus fits four parameters
(plus a phase) for blocks of data that typically extend over 10--60\,sec in
time and up to 48\,MHz in frequency.

In addition it is found, not unexpectedly, that differential Faraday rotation
between the stations is generally significant at lower frequencies and on long
baselines. This means that the XX and YY correlations are not sufficient to
determine Stokes I, because parts of the signal are shifted to the mixed
correlations XY and YX. Finally the imaging on long baselines is much more
difficult than for the Dutch array alone, because the $uv$
coverage is much more sparse.

\begin{figure}[hb]
\raisebox{0.01\textwidth}{\includegraphics[width=0.45\textwidth]{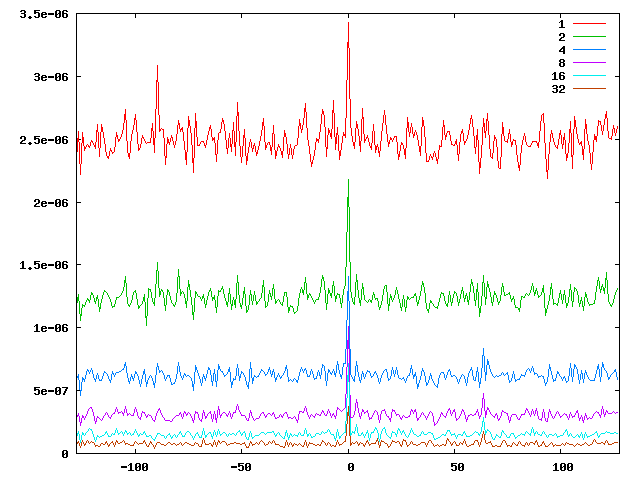}} \hfill
\includegraphics[width=0.49\textwidth]{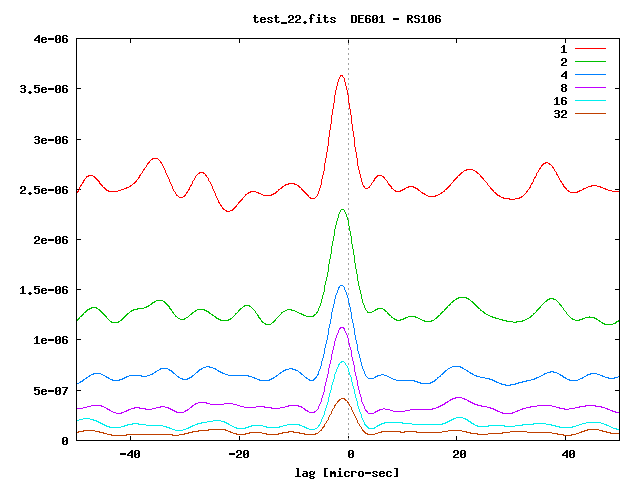}
\caption{Delay spectra on a baseline between the Netherlands and
  Effelsberg. The different colours denote different coherent averaging times
  (labels in units of the integration time of 5\,sec).
 The second plot with higher delay resolution shows that the peak
  is slightly offset from 0.}
\label{fig:fringes delay}
\end{figure}

\section{First long-baseline fringes}

Shortly after others had seen hints of short periods of consistent phases on
LBA observations of 3C196 that included the station Effelsberg, I started a
more systematic search for 
fringes in the parameter space of delay and rate.
For the first attempt in August 2009, data of one subband were coherently
averaged over short intervals, then Fourier transformed on the frequency axis
to obtain delay spectra, and the power was averaged (incoherently) over one
hour to increase the $S/N$. Results are shown in Fig.~\ref{fig:fringes delay}.

The clear peak in the delay spectra near a lag of 0 did not convince everybody
at that time,
because other effects may mimic real fringes caused by a sky signal,
particularly if the peak is at exactly zero delay. Plotting the spectra with
higher resolution revealed that the peak is in fact offset by about one
$\mu$sec.

\begin{figure}[h]
\includegraphics[width=0.48\textwidth]{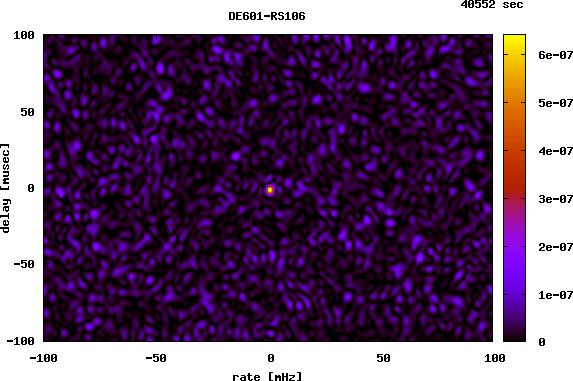} \hfill
\includegraphics[width=0.49\textwidth]{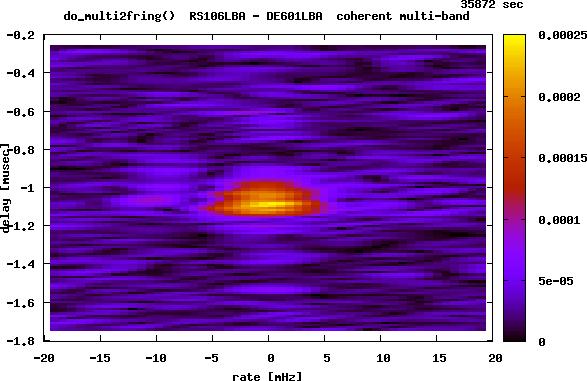}
\caption{Fringe-rate/delay spectra for a short period of observations of 3C196
on a baseline between the Netherlands and Effelsberg. Left: single-band with
low resolution on the vertical delay axis, right: coherent combination of many
bands with much higher resolution in delay. The strange shape of the peak is
partly caused by not including dispersive effects in this plot.
Animated versions are available at \myurl{http://www.astro.uni-bonn.de/~wucknitz/download/2009_dwingeloo_lofar_status_fring4a.gif} and \myurl{http://www.astro.uni-bonn.de/~wucknitz/download/2009_dwingeloo_lofar_status_msfring1_coh.gif}.
}
\label{fig:fringes 2d}
\end{figure}

A more systematic analysis followed in which the time-dependence was
explicitly taken into account by scanning a two-dimensional (and finally
four-dimensional) parameter space of 
delay and rate. See Fig.~\ref{fig:fringes 2d} for the first single-band and
multi-band results, again for observations of 3C196.

This analysis started on the short baselines, where the strongest fringes were
surprisingly found for delays and rates far from zero and of the same order of
magnitude as the total geometric delays and rates. It was first thought that
the delay tracking did not work properly, but then I found that the
stronger fringes are coming from other bright sources like Cygnus~A or
Cassiopeia~A. Their signal, leaking in through high-order sidelobes of the
station beam, can be
stronger than the target signal even if they are more than 90\deg\ away from
the station beam centre.

\begin{figure}
\center
\includegraphics[width=0.75\textwidth]{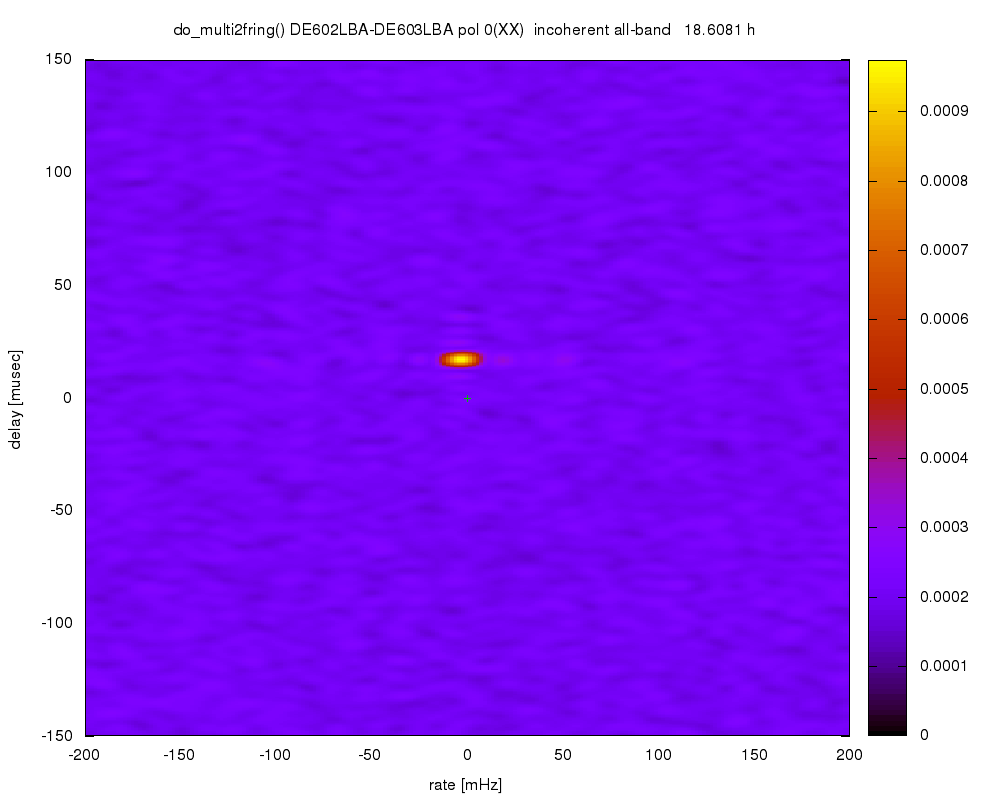}
\caption{First fringes on the baseline Tautenburg--Unterweilenbach (January
  2010). The delay of about $17\,\mu$sec is caused by a clock offset in the
  Tautenburg station that resulted from an incorrect station position in the
  GPS configuration. This problem was fixed later.
In the movie
(\myurl{http://www.astro.uni-bonn.de/~wucknitz/download/2010_feb17_lofar_status_anim.mp4})
we sometimes see other peaks moving around slowly. They are caused by other
sources in the field.}
\label{fig:fringes DE-DE}
\end{figure}

The next two German stations coming online were Tautenburg and Unterweilenbach
with first LBA observations in January 2010. Shortly thereafter I detected
fringes to Tautenburg, then to Unterweilenbach and finally even between pairs
of German stations. The first fringes between Tautenburg and Unterweilenbach
are shown in Fig.~\ref{fig:fringes DE-DE}.

\begin{figure}
\center
\includegraphics[width=0.65\textwidth]{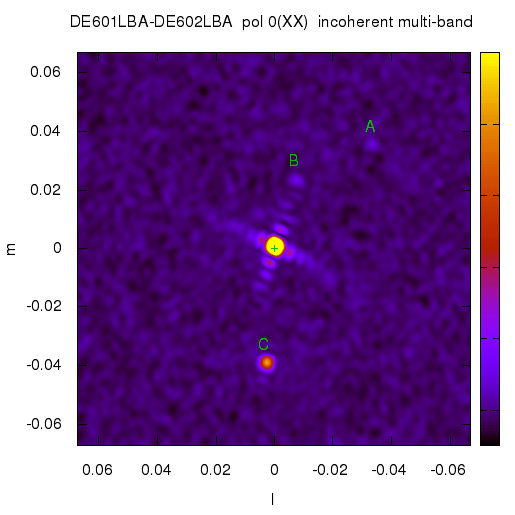}
\caption{Delay/fringe-rate map around 3C196 created from delay/rate spectra of
the long baseline, Effelsberg--Unterweilenbach. The bright source in the
centre is 3C196 with about 140\,Jy. The other three sources are listed in the
VLSS with fluxes of 19/6/17\,Jy for A/B/C at 74\,MHz,
respectively. Coordinates are $l,m$ in radians, 0.04 corresponds to about
2\fdg3. This map is also the JIVE/ASTRON
  picture of the day of
\protect\href{http://www.astron.nl/dailyimage/main.php?date=20100303}{3rd March 2010}.
}
\label{fig:delay/rate map}
\end{figure}

Applying the elementary
geometry of the Earth rotation, I transformed delay/rate values to sky
position and in this way (after averaging over a short period) produced a
delay/fringe-rate map of the field around 3C196 using data from only one
baseline (see Fig.~\protect\ref{fig:delay/rate map}). Besides the target
3C196, three other sources can be seen more or less clearly.

\begin{figure}
\center
\includegraphics[width=0.82\textwidth]{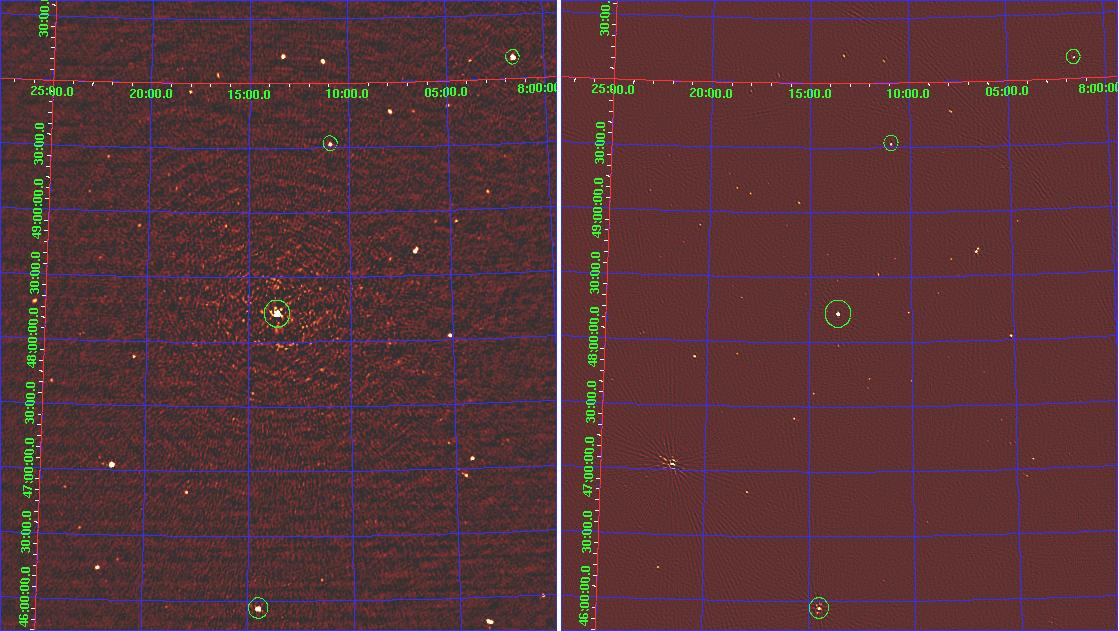}
\caption{VLSS map (left) and LOFAR LBA map (right) of the field around
  3C196. Only Dutch baselines were used in the imaging. 3C196 in the centre
  was used as the (only) calibrator. The three sources A/B/C from
  Fig.~\protect\ref{fig:delay/rate map} and 3C196 in the centre are marked
  with circles. We see that, as expected, the calibration is much better close
  to the centre of the field than further away.}
\label{fig:widefield map}
\end{figure}

For comparison, a wide-field synthesis map was produced using only Dutch
baselines. Fig.~\ref{fig:widefield map} shows the result  and a VLSS map of
the same field.

\section{Results from fringe-analysis}

The most important result is the clear detection of fringes in long-baseline
LBA observations. In addition a number of technical issues were found in the
course of this analysis.
Among them are the relatively large clock offsets in German stations, some
confusion about the labelling of LBA polarisations under certain circumstances,
a strong 8\,MHz ripple that resulted from accidentally disabling the
cable-length compensation in the beam forming, the detection of strong
differential Faraday rotation and an independent detection of a resonance near
63\,MHz. 

Most of these problems have been fixed later. The independent long-baseline
analysis contributed significantly to that process, even though this was not
the primary intention of my efforts.

\section{First long-baseline maps: LBA}

For the first serious long-baseline imaging attempt, a 6\,h observation on the target 3C196
on 12/13th Feb 2010 (data set D2010\_16704) was used. Five Dutch and three
German stations (Effelsberg, Unterweilenbach and Tautenburg) produced a wide
range of baselines lengths. Only a fraction of the 160 subbands (30--80\,MHz
with small gaps) showed a strong signal because of the aforementioned 8\,MHz
ripple. In the first attempts, only the good subbands were used, later the
weaker ones were included, too.

In the analysis I first corrected for the $1\,\mu$sec and $17\,\mu$sec clock
offsets in Effelsberg and Tautenburg, applied some flagging, then converted
the XX/XY/YX/YY polarisation combinations to circular RR/RL/LR/LL in order to
avoid differential Faraday rotation, averaged somewhat in time and frequency,
and then finally imaged and self-calibrated RR and LL in \aips\ and \difmap. 
For the final maps produced with \difmap, Gaussian components with variable
spectral indices were fitted to the data in order to correct the MFS (multi
frequency synthesis) imaging for spectral index variations.

\begin{figure}[ht]
\center
\includegraphics[width=0.46\textwidth]{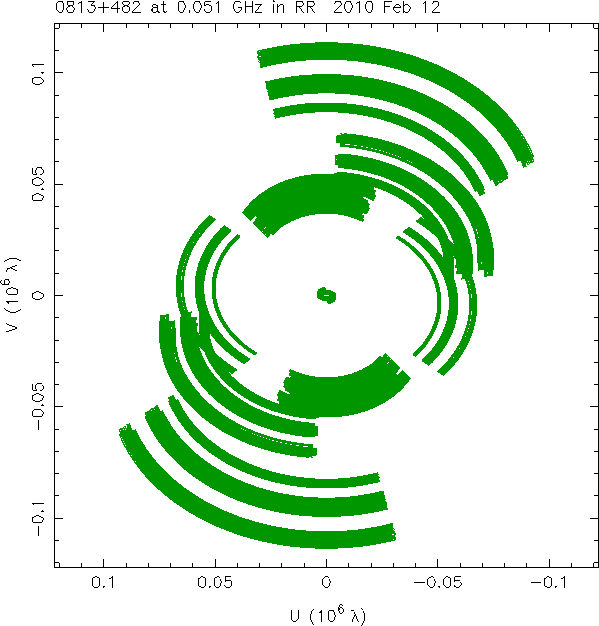}\hfill
\includegraphics[width=0.46\textwidth]{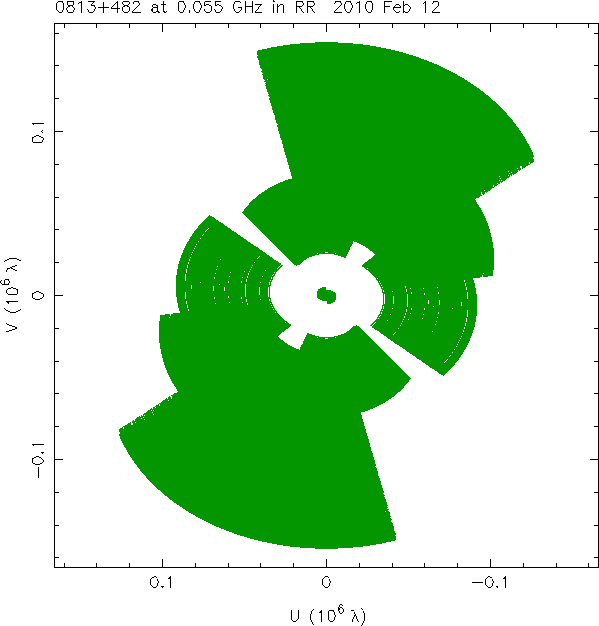}
\caption{$uv$ coverage of the long-baseline LBA observations. Left: only the
  `good' subbands, right: the entire frequency range. The Dutch baselines form
  the compact small region in the centre. All other baselines are combinations
  of/with German stations.}
\label{fig:LBA uv cov}
\end{figure}

\begin{figure}
\includegraphics[height=0.43\textwidth]{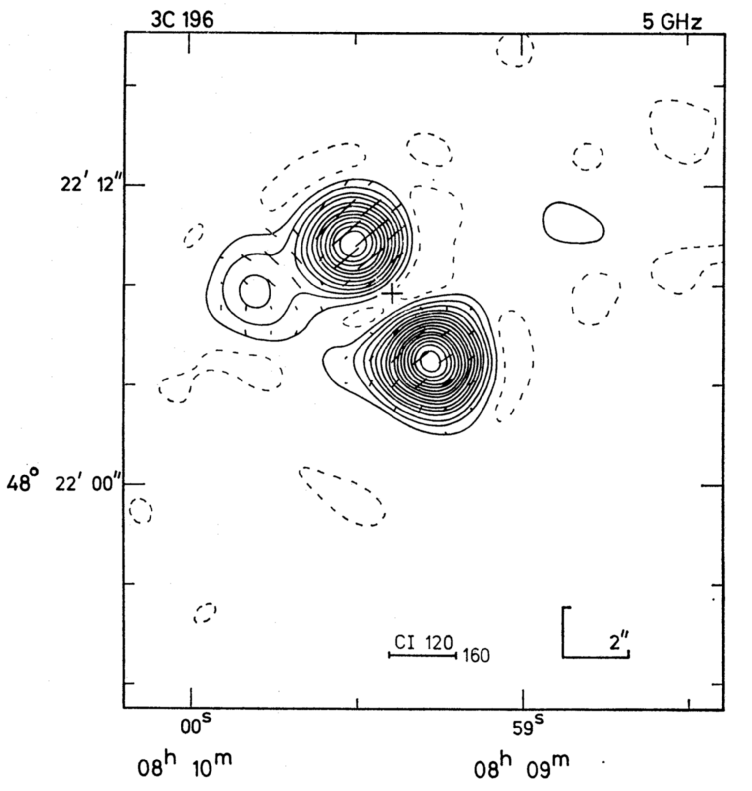}\hfill
\includegraphics[height=0.42\textwidth]{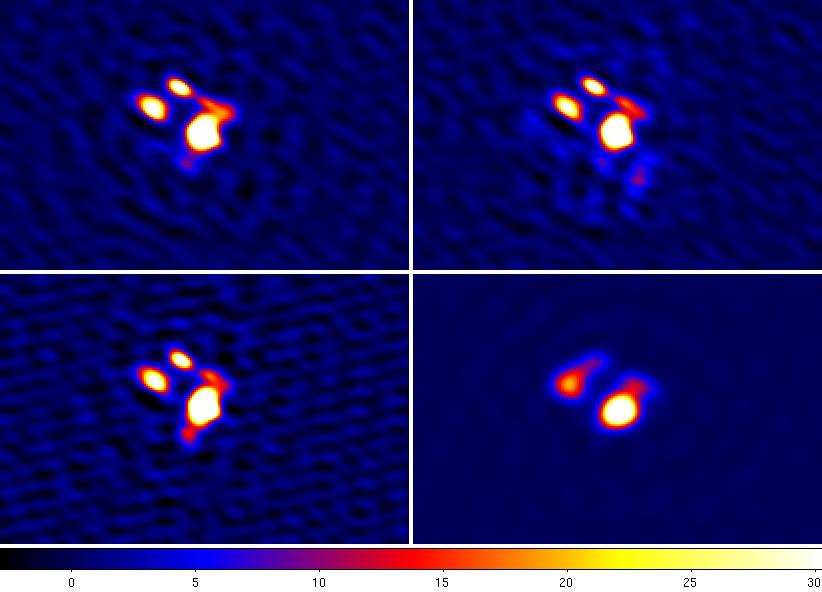}
\caption{Left: 3C196 at 5\,GHz observed with the Cambridge 5\,km array
  \citep[from][]{pooley74}, component separation ca.\ $5\farcs5$, right: four early attempts to image the same
  source from LOFAR LBA data using slightly different calibration
  strategies. All four panels show basically the same structures.
 The lower right panel was produced form a different data set
  with only one German station (Effelsberg). In this case the long-baseline
  information came from the amplitudes alone, because no closure phases are
  available with effectively only one long baseline. The amplitudes were
  stable during the observation.
}
\label{fig:LBA images 1}
\end{figure}

\begin{figure}
\center
\includegraphics[width=0.7\textwidth]{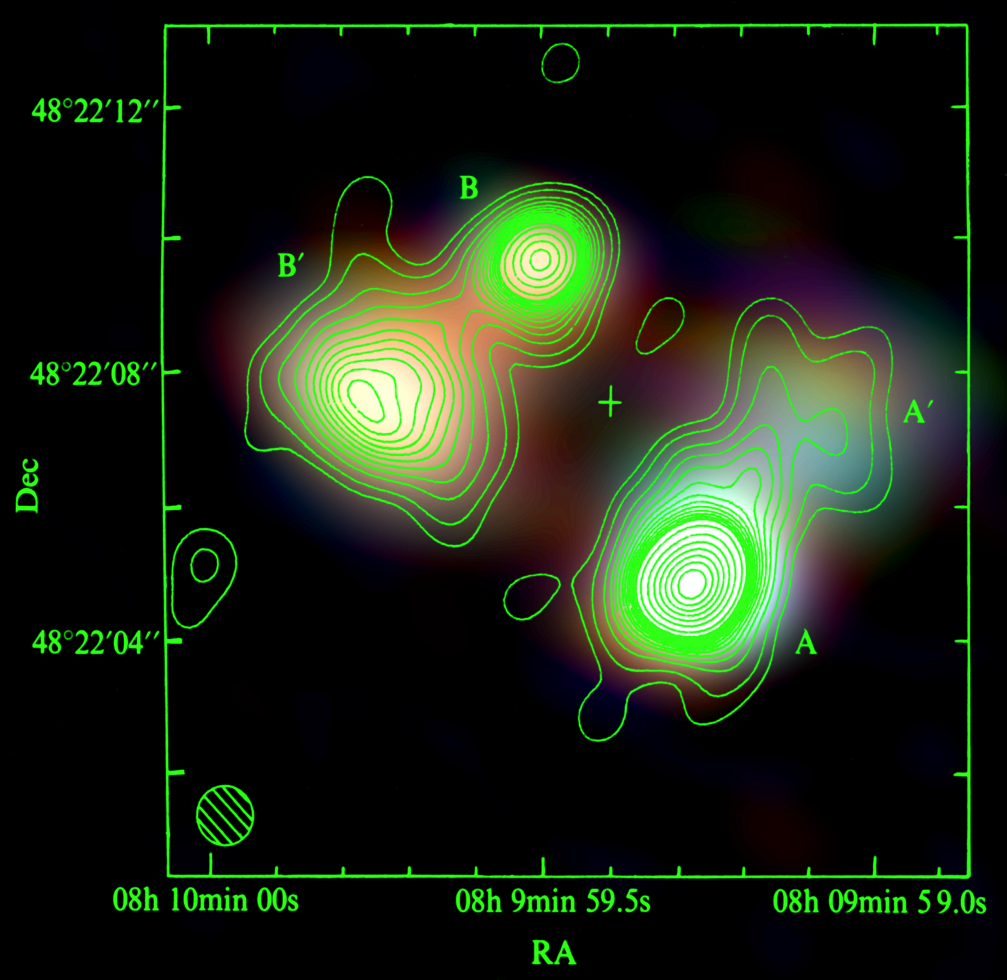}
\caption{Comparison of the LOFAR image of 3C196 (background image) with the
  408\,MHz observations (contours from \citet{lonsdale80}). The structures
  match up surprisingly well, given the factor of 8 in frequency. 
  Note that the self-calibration process started conservatively with a
  point-source model and   continued without prior knowledge of the 408\,MHz
  details.} 
\label{fig:map overlay}
\end{figure}

The $uv$ coverage is shown in Fig.~\ref{fig:LBA uv cov}.
The expected appearance of the source and the very first LOFAR images are shown in
Fig.~\ref{fig:LBA images 1}. To my surprise, about four instead of the
expected two components showed up in the images. However, the inspection of
408\,MHz MTRLI (MERLIN) observations revealed a very similar structure (see
Fig.~\ref{fig:map overlay}).

\begin{figure}
\includegraphics[width=\textwidth]{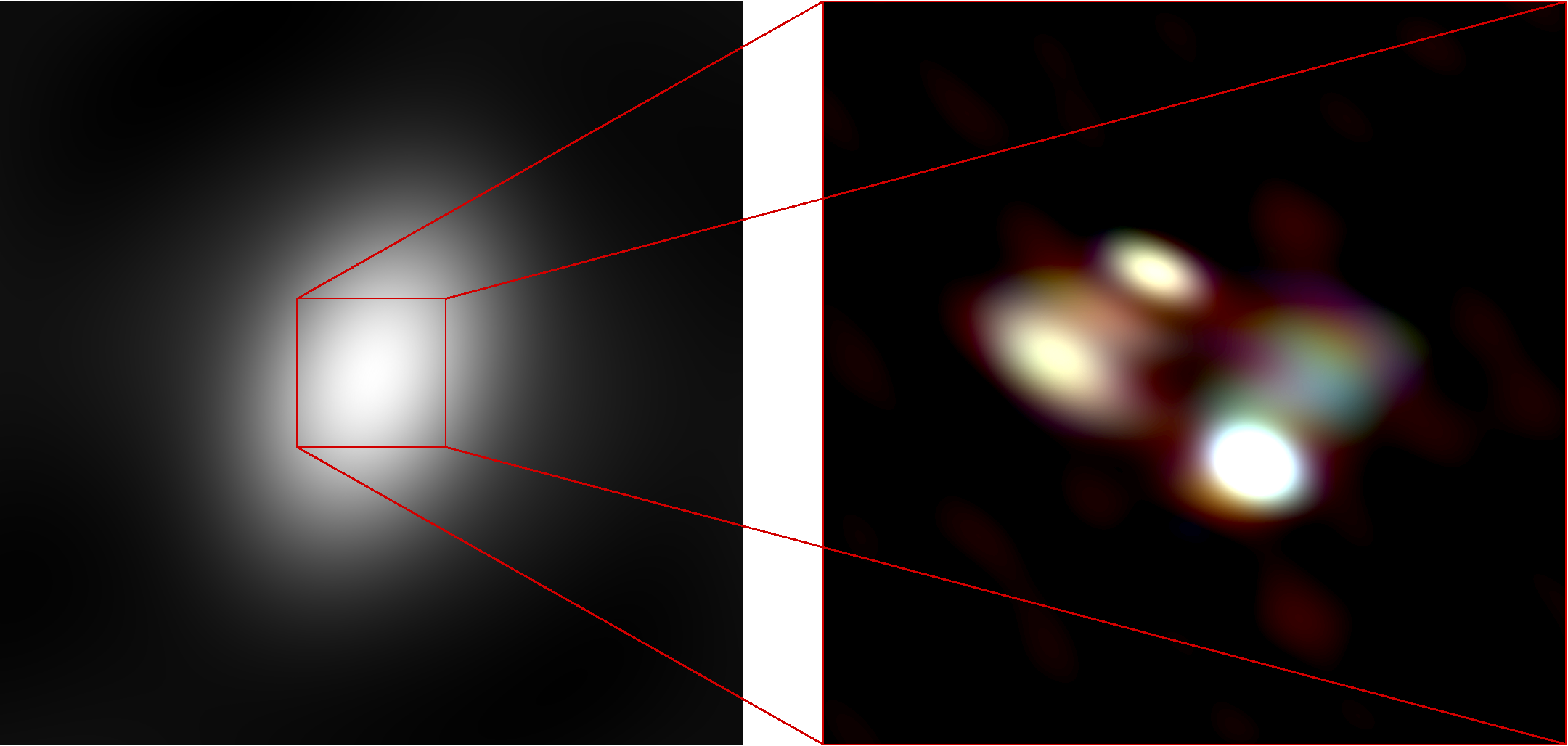}
\caption{Left: Map of 3C196 using only the Dutch baselines. The resolution of
  $35''\times 22''$ is not sufficient to resolve any details. Right: Blowup
  using all the long baselines. With a resolution of $1\farcs5 \times
  0\farcs9$, the structures of the target are finally revealed. The colours
  are chosen to resemble what the human eye would see if it were sensitive to
  radiation at a wavelength ten million times larger than visible light. 
This is also the JIVE/ASTRON
  picture of the day of
\protect\href{http://www.astron.nl/dailyimage/main.php?date=20100601}{1st June 2010}.
}
\label{fig:LBA map}
\end{figure}

The comparison with the MERLIN images gave more confidence in the reliability
of the imaged structures. The process was repeated and improved to produce the
currently best long-baseline LBA LOFAR image shown in Fig.~\ref{fig:LBA map}.
The direct comparison with the 408\,MHz image is shown in Fig.~\ref{fig:map
  overlay}.

These maps have the highest resolution that has \emph{ever} been achieved in
this frequency range so far. They are a great step forward for low-frequency
VLBI. 

\section{First long-baseline maps: HBA}

Progress with the HBA is still lagging behind. Shortly after the first
long-baseline HBA observations, fringes could be confirmed, but they turned
out to be much weaker than expected. The imaging was correspondingly intricate
and has not produced satisfying results so far.

For the first image, observation L2010\_07608 of 3C196 was used (12\,h on 22nd
May 2010). The frequency range was 131--155\,MHz in 120 subbands. 7 Dutch and
only two German stations (Effelsberg and Tautenburg) were used. I had to
correct for the clock offset in Effelsberg and for an offset of $8\,\mu$sec in
the superterp stations in the core in order to allow averaging in frequency.
The imaging and self-calibration was done in \difmap\ and used the YY
correlation.
Most of the time the $S/N$ on the long baselines was very low, particularly on
the baseline Effelsberg--Tautenburg. The reason for this is not fully
understood. Partially it is due to the source structure, but more important
seems to be the (then still missing) station calibration. Some inconsistencies
between blocks of subbands made the averaging in frequency difficult.

\begin{figure}
\includegraphics[height=0.49\textwidth]{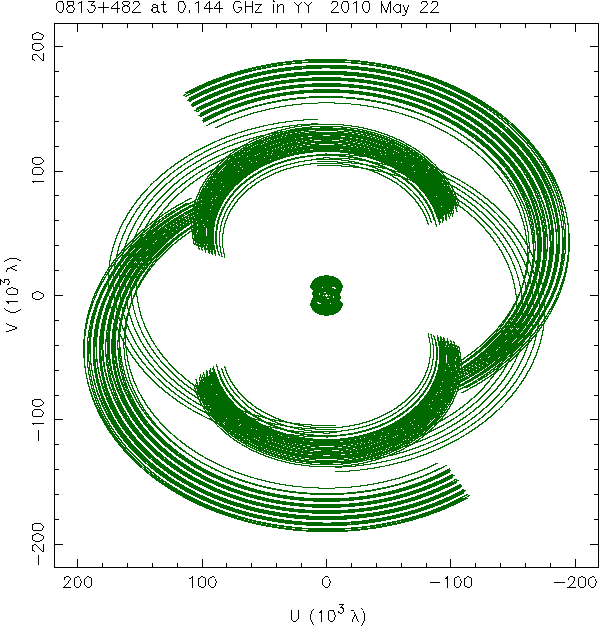}\hfill
\includegraphics[height=0.49\textwidth]{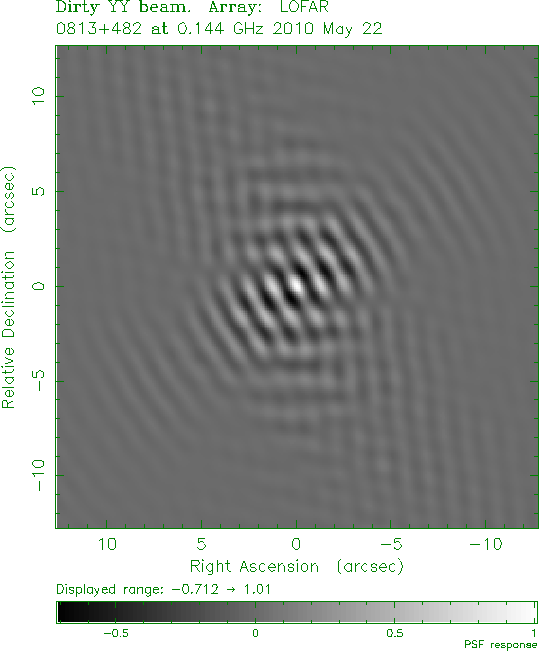}
\caption{Left: $uv$ coverage, right: calibrated dirty beam of the HBA
  observations of 
  3C196. The strong sidelobes make the imaging process very unstable and
  difficult.}
\label{fig:HBA uv beam}
\end{figure}

\begin{figure}
\center
\includegraphics[width=0.82\textwidth]{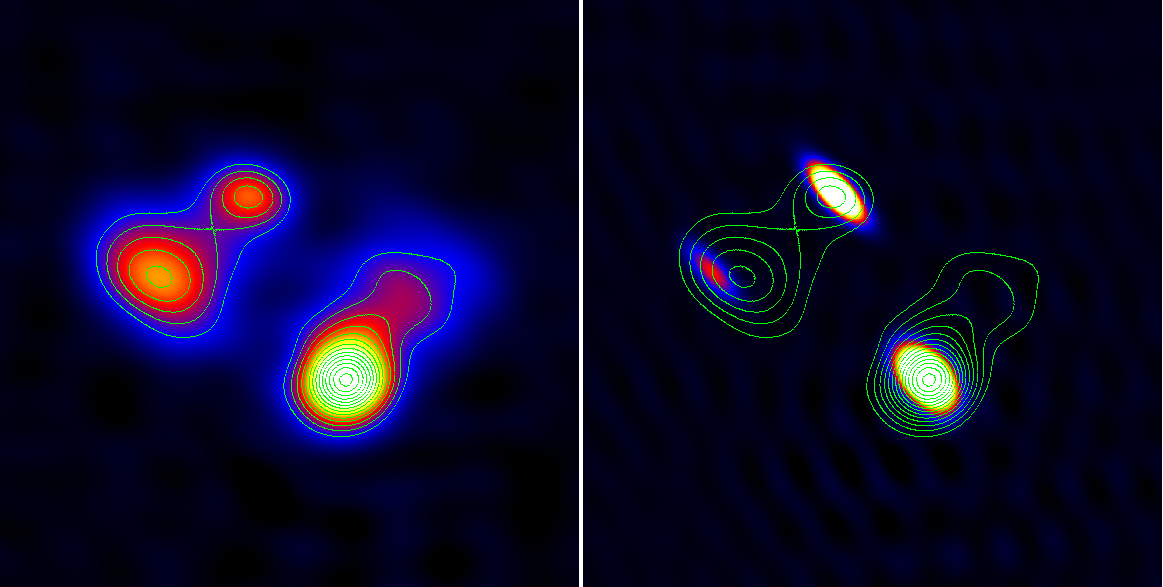}
\caption{Left: LBA map described before, right: preliminary HBA map with
  superposed LBA contours. Even though the details are not 100\,\% reliable
  yet, it looks as if we only see the brightest and most compact regions
  in three of the components. They do not coincide with the centroids of the
  LBA emission. This is not entirely surprising, because all components seen
  are lobes with unresolved hot spots. The core of the AGN is not seen at radio wavelengths.}
\label{fig:HBA map}
\end{figure}

The $uv$ coverage and dirty beam are shown in Fig.~\ref{fig:HBA uv beam}, the
very preliminary map in Fig.~\ref{fig:HBA map}. So far we only see compact
parts of three of the components at HBA frequencies.

\section{Acknowledgements}

I would like to thank many colleagues in the LOFAR team whose work made these
results possible. In particular the contributions of the following people were
directly related to the results presented here. From ASTRON:
Ashish Asgekar,
Michiel Brentjens,
Andr\'e Gunst,
George Heald,
John McKean,
Menno Norden,
Antonis Polatidis and
Stefan Wijnholds. From the German stations: James Anderson
(Effelsberg), Annette Haas and Matthias Hoeft (Tautenburg).

\section{Summary}

This contribution proves that the long baselines of LOFAR actually work. They
produce fringes, they can be calibrated, and they can be used to produce
high-resolution images of the low-frequency radio sky.
The highest-resolution image in the frequency range 30--80\,MHz has been
produced from only three international (plus five Dutch) stations, a small
part of the final array. Fringe-fitting methods are required to determine and
correct dispersive and non-dispersive delays in order to allow averaging in
frequency. Phase changes with time turn out to be not much worse on long
baselines than on short ones. Beyond about 30\,km, the ionospheric patches are
more or less independent of each other, so that variations do not increase
with baseline length anymore. Even though the delays are larger, they can be
calibrated because of their stability.

In order to analyse these data, new analysis techniques had to be developed
and implemented. The results presented here were achieved with software
written particularly for this purpose together with the standard packages
\aips\ and \difmap.

The original presentation of the conference talk together with a number of
fringe movies is available from the author's homepage at
\myurl{http://www.astro.uni-bonn.de/~wucknitz/publications/pub.php?2010_iskaf_assen_lofar}.

\bibliographystyle{proceedings}
\bibliography{proceedings}

\end{document}